\documentclass[
aps,prl,
reprint,
superscriptaddress,
]{revtex4-1}

\usepackage{graphicx}
\usepackage{dcolumn}
\usepackage{bm}
\usepackage{xcolor}
\usepackage{mathrsfs,amsmath}
\usepackage[colorlinks = true,
            linkcolor = blue,
            urlcolor  = blue,
            citecolor = blue,
            anchorcolor = blue
            ]{hyperref}
\newcommand\rxout{\bgroup\markoverwith{\textcolor{red}{\rule[.5ex]{2pt}{.6pt}}}\ULon}

\begin{document}

\title{Omnidirectional elastic wave attenuation via an isotoxal-star-based auxetic micro-lattice}
\author{Nikhil JRK Gerard}
\affiliation{Department of Mechanical and Aerospace Engineering, North Carolina State University, Raleigh, North Carolina, USA 27695}
\author{Mourad Oudich}%
 \email{moudich@ncsu.edu}
 \affiliation{Department of Mechanical and Aerospace Engineering, North Carolina State University, Raleigh, North Carolina, USA 27695}
\affiliation{Univerist\'e de Lorraine, CNRS, Institut Jean Lamour, F-54000 Nancy, France}

\author{Yun Jing}
 \email{yjing2@ncsu.edu}
 \affiliation{Department of Mechanical and Aerospace Engineering, North Carolina State University, Raleigh, North Carolina, USA 27695}

\date{\today}

\begin{abstract}This paper introduces a micro-lattice based metamaterial for low frequency wide-band vibration attenuation, that is enabled by engineering the metamaterial's building blocks to induce local resonance bandgaps for elastic waves in all directions of propagation. The transmission rate through the proposed structure is examined and strong wave attenuation is demonstrated for a remarkably small number of unit cells. Additionally, it is shown that the bandgaps are tailorable via the geometrical parameters and can be leveraged to design a hybrid metamaterial with an extremely wide bandgap. Alongside being thin, lightweight, and capable of attenuating elastic waves in all directions, the proposed material also possesses the second order functionality of exhibiting a negative Poisson's ratio and can pave the way for identifying exotic functional materials.  
\end{abstract}

\maketitle

\begin{figure*}
\includegraphics{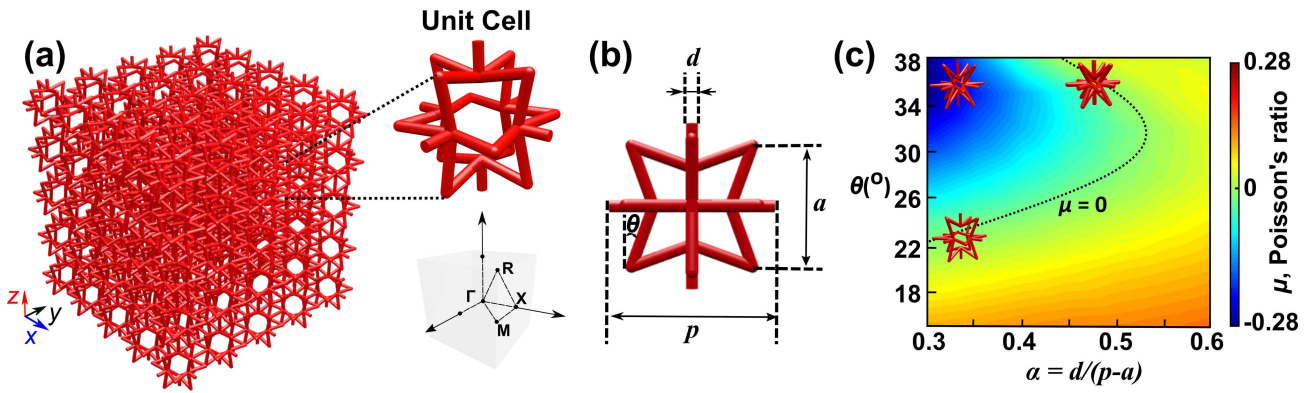}
\caption{\label{FirstFigure}(a) Micro-lattice-based metamaterial composed of periodic auxetic building blocks(top) and the directions of the irreducible Brillouin Zone (IBZ) for the simple cubic unit cell(bottom). (b) Dimensions of the unit cell - $p$ is the periodicity and $a$ is the distance between the verticies of the isotoxal squares. $\theta$ is the angle between by the outer edges of the isotoxal square stars and the vertical direction as shown. (c) Poisson's ratio, $\mu$, for different values of $\theta$ and $\alpha = d/(p-a)$. Insets shows the geometry of the unit cell for the largest negative value of $\mu$ and for when $\mu = 0$ respectively.}
\end{figure*}
The search for materials with exceptional physical properties has accompanied and steered technological progress for the past two centuries. For example, lightweight materials are extremely desirable since low density helps increase the longevity of a material that is subjected to successive and repetitive mechanical loading. It is not straightforward however, to create a material that is both lightweight and mechanically strong - two properties that are highly favorable for aerospace and automotive applications and for energy saving in general. Interestingly though, nature provides us with several materials where both these characteristics co-exist. Wood \cite{Song2018}, bones and exoskeletons \cite{Meyers2008} are well known examples of such lightweight biological materials which possess remarkable mechanical responses, due to their highly ordered hierarchical fibrous structures at the micro- and nanoscales. Drawing inspiration from these biomaterials, recent researchers have employed contemporary fabrication techniques to mimic their microstructure and put forward materials with ultra-low density and exceptional mechanical behavior. Graphene based elastomers for super elasticity\cite{Qiu2012}, carbon nanotube foams \cite{Worsley2009} and ultra-light nano-porous silica (aerogels) for high compressibility \cite{Kucheyev2012} are a few examples of the several propositions in this regard. Furthermore, the advent of more state-of-the-art additive manufacturing techniques have enabled the fabrication of architected materials with precise micrometric and even nanometric geometrical features, with multiple orders of hierarchy \cite{Zheng2014,Zheng2016}. This thus gave birth to low-density three-dimensional (3D) materials made up of lattice structures that are architected for maximizing recoverability under compression \cite{Schaedler2011,Torrents2012}, super-elastic tensile behavior \cite{Zheng2014,Zheng2016} or decoupling density and mechanical behavior \cite{Hedayati2017}.

Alongside such works, auxetic materials which display negative Poisson's ratio have emerged \cite{Lakes1987,Evans1991}. When subjected to a tensile (or compressive) load, the volume of such materials would counter-intuitively increase (or decrease). This intriguing behavior brought about extensive works that later showed that two-dimensional (2D) \cite{Grima2005,AlvarezElipe2012} and 3D \cite{Hengsbach2014,ShokriRad2014,Imbalzano2016,Saxena2016,Chen2018,Ren2018} auxetic materials possess great potential as lightweight impact energy absorbers \cite{Imbalzano2017,Xie2019}. Efforts have also revealed that in addition to the conventional auxetic geometries \cite{Lakes1987}, auxetic behavior can also be induced via chirality \cite{Spadoni2011,Dirrenberger2013,Ha2016}, rotating units \cite{Grima2000,Attard2008} and random or entangled topologies \cite{Grima2016,Rodney2016}. Thus, one of the recent trends in this area inclines towards micro-lattice-based materials where auxetic structures serve as fundamental building blocks. Coupling this design strategy with multi-material additive manufacturing has also enabled giant tunable values of negative Poisson’s ratio \cite{Chen2018}, a characteristic that is far from what one would observe in nature. It is thus evident that the concept of metamaterials that was bound to electromagnetism and acoustics, has now been extended to embrace this new class of 3D mechanical materials.

Such complete control over material and geometrical features at the micro-scale is also enticing from the perspective of wave propagation. Over the past two decades, phononic crystals and acoustic metamaterials have demonstrated that periodic geometries can be engineered to possess desirable bandgaps or frequency ranges of no propagation. Most research here, however, has been centered around 2D geometries, since the initially proposed 3D designs that employ cubic structures of cores embedded in a matrix, are too bulky for real world applications and cumbersome to fabricate \cite{Liu2000,Yang2004,Cheng2006,Lu2017,Lucklum2018}. Moreover, conventional 3D printing techniques can quite conveniently realize elastic metamaterials that are designed for 1D \cite{Matlack2016}, 2D \cite{Bilal2017} and 3D \cite{Bilal2018,Taniker2015} wave attenuation. Interestingly enough, however, auxetic geometries have come to light in this context as well, since they facilitate relatively wide bandgaps, while maintaining the desired weight. Although multiple studies have investigated the elastic wave dispersion through 2D auxetic structures \cite{Meng2015,Mukherjee2016,Bacigalupo2016}, works that explore the same in 3D still remain scarce \cite{Brunet2015,Krodel2014,DAlessandro2018}. Additionally, the  existing 3D designs employ masses to introduce bandgaps or to increase its width, which once again increases the material weight making it bulky for scenarios where size is paramount. The dawn of micro-lattice-based metamaterials, however, could serve as a prospect to overcome these limitations. In this paper, we put forth a 3D auxetic micro-lattice that is equipped with tailorable sub-wavelength bandgaps for elastic waves. The proposed elastic metamaterial shields waves in all directions and over a wide range of frequencies, alongside exhibiting a static negative Poisson's ratio. Such a thin lightweight material with dual functionality could thus pave way for a variety of  unconventional wave-based devices, alongside bolstering the interest for futuristic multi-functional materials. 

\begin{figure}
\includegraphics{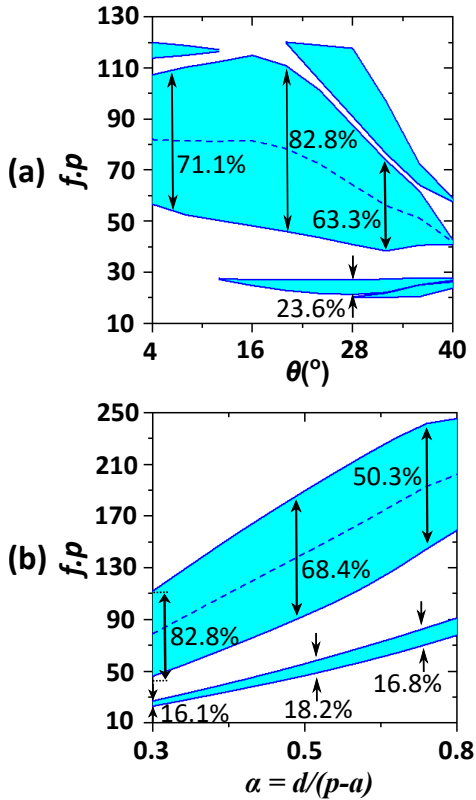}
\caption{\label{ThirdFigure}The evolution of the bandgaps (shaded cyan regions) along the normalize frequency ($f.p$) spectrum as a function of $\theta$ (a) and $\alpha = d/(p-a)$ (b).}
\end{figure}

The proposed metamaterial is made up of simple cubic auxetic unit cells, as shown in the Fig. \ref{FirstFigure}(a). All the strut elements of the lattice are made up of solid cylindrical rods of the same diameter and intrinsic material. However, different aspect ratios are employed for the inner and the outer rods respectively. 24 inner rods make up the 3 mutually perpendicular isotoxal square stars in the $x$,$y$ and $z$ planes. The 6 outer rods have a slightly different aspect ratio and serve to connect each unit cell with its adjacent neighbors. The defining dimensions as shown in Fig. \ref{FirstFigure}(b), are thus $p$ and $a$, which denote the periodicity and the distance between the vertices of the squares respectively. In the calculations that follow, $a = 0.8p$ and the diameter of the cylindrical rods, ‘$d$’, is chosen such that the parameter $\alpha$ = $d/(p-a)$ is less than 1. This is done so as to ensure that the vertices of the isotoxal stars do not overlap. The two geometrical parameters that are treated as variables are, therefore, the angle $\theta$ (in Fig. \ref{FirstFigure}(b)) and $\alpha$ which controls the strut diameter while maintaining the specified constraint. Furthermore, the intrinsic material properties utilized are that of 1,6-Hexanediol diacrylate (HDDA), a well-known polymer in the context of micro-stereolithography \cite{Zheng2014,Zheng2016,Gerard2019} – the additive manufacturing technique that can enable the realization of such a micro-lattice. HDDA has a Young’s Modulus, E, of 512 MPa, a density, $\rho$, of 1.1 g/cm$^{3}$ and an intrinsic Poisson’s ratio, $\nu$, of 0.3. To verify that the proposed material is auxetic, numerical simulations were first carried out by means of a stationary study on the solid mechanics module on COMSOL Multiphysics 5.3(a), a commercial finite element package. A known longitudinal displacement was prescribed on one face of a finite sample made up of 3x3x3 unit cells, and the resultant lateral displacement was extracted by the method described in Ref .\citep{Imbalzano2017}, to estimate the Poisson’s ratio of the bulk material. Figure \ref{FirstFigure}(c) shows the value of Poisson’s ratio for different values of $\theta$ and $\alpha$. As can be seen, the metamaterial exhibits auxeticity when the parameters lie in the region above the dashed curve, which corresponds to $\mu$ = 0. The Poisson’s ratio can be further lowered by decreasing (increasing) the values of $\alpha$ ($\theta$) and can reach a minimum of -0.28 for $\theta$ = 35$^\circ$ and $\alpha$ = 0.3. 
\begin{figure*}
\includegraphics{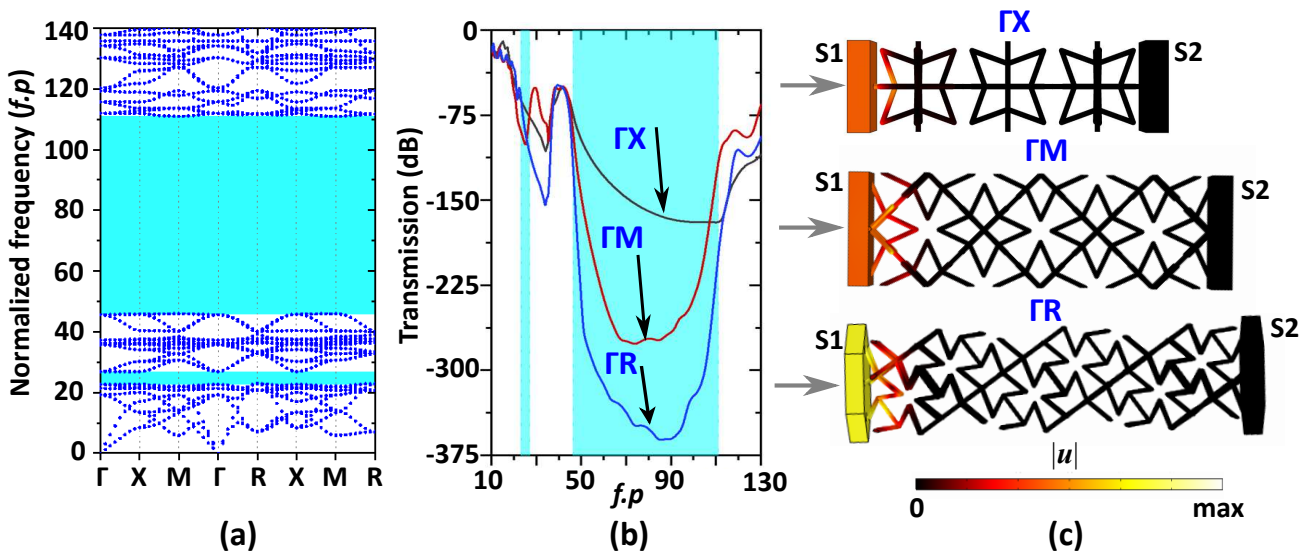}
\caption{\label{SecondFigure}(a) Bandstructure for the simple cubic auxetic unit cell along the different directions of the IBZ for $\theta = 20^\circ$ and $\alpha = 0.3$. (b) Elastic wave transmission along three units in the three directions $\Gamma$X, $\Gamma$M and $\Gamma$R, with an istoropic loss factor of 0.05 (The cyan regions in (a) and (b) indicate the bandgaps). (c) Total displacement field amplitude $|\textbf{\textit{u}}|= \sqrt{|u_{x}|^{2}+|u_{y}|^{2}+|u_{z}|^{2}}$ for $\Gamma$X, $\Gamma$M and $\Gamma$R at the normlized frequency $f.p = 85$.}
\end{figure*}
The elastic wave dispersion through the micro-lattice was then examined via an eigenfrequency study on a single unit cell with Floquet periodic boundary conditions. The bandstructure was calculated along the principle directions of the irreducible Brillouin zone (IBZ) described in Fig. \ref{FirstFigure}(a). Figure \ref{SecondFigure} shows the bandstructure when $\theta$ = 20$^\circ$ and $\alpha$ = 0.3. Two complete bandgaps of widths 16.1\% and 82.8\% can be seen (regions in shaded cyan color) in the normalized frequency ($f.p$) ranges of 22.9 to 26.9 and 45.95 to 110.95. This implies that if the unit cell has a period of 2 cm, the second bandgap would exist between 2.3 kHz and 5.55 kHz. To further corroborate the existence of such bandgaps and to evaluate its attenuation performance, frequency domain transmission simulations were performed for a fixed number of unit cells along different directions of propagation. The unit cells were bounded by two thin plates of thickness 0.25$p$ on either side and periodic boundary conditions were assigned for the other sides of the sample. Additionally, a realistic amount of  viscoelastic loss (isotropic loss factor  = 0.05), was also included in the simulation. A harmonic boundary load was applied to the surface of the first plate (S1) and the average of the total displacement field along the surface of the second plate (S2) was extracted, to calculate the transmission coefficient Fig. \ref{SecondFigure}(c). This was done for three samples to examine the propagation along the different directions of the defined IBZ. It must thus be noted that the $\Gamma$X, $\Gamma$R and $\Gamma$M samples considered here have periodicities of $p$, $\sqrt{2}p$ and $\sqrt{3}p$ respectively, and the wave transmission was studied over 3 periods for each of these cases. The transmission curves, shown in Fig. \ref{SecondFigure}(b), clearly exhibit sharp decays at the normalized frequency ranges between 22.9 and 26.9 and once again between 45.95 to 110, in accordance with the two bandgaps previously depicted in Fig. \ref{SecondFigure}(a). When a wave in these frequency ranges propagates through the micro-lattice, a local resonance occurs due to the trampoline-like nature of the geometry \cite{Bilal2017,Bilal2013}, thereby entirely reflecting the wave. As one may point out, at lower frequencies, the transmission curves in the three different directions display different decay regions with overlapping bandgaps. This corresponds to partial bandgaps for longitudinal waves in each of the directions since the excitation here is perpendicular to S1. Meanwhile, in the frequency range between 45.95 and 110.95, a very high attenuation (more than 150dB) is seen for all three cases. The decay is subsequently higher in the cases of $\Gamma$M ($\sim$ 275 dB) and $\Gamma$R ($\sim$ 375 dB), as the sample is proportionally longer due to the slightly larger value of unit periodicity. This can also be depicted in Fig. \ref{ThirdFigure}(c), which shows the displacement field amplitude at the normalized frequency, $f.p$ = 85 inside the bandgap. As can be quite clearly seen, the resonance occurs at the interface between the outer and the inner rods of the auxetic geometry and the wave hence only propagates through half of the first unit in all cases. Such a high wave attenuation with only three units makes this design extremely advantageous for low frequency vibration shielding. For instance, a sample of thickness 3.5 cm ($p$ = 1 cm) would possess a wide bandgap between 4.6 kHz and 11.1 kHz.  
\begin{figure*}
\includegraphics{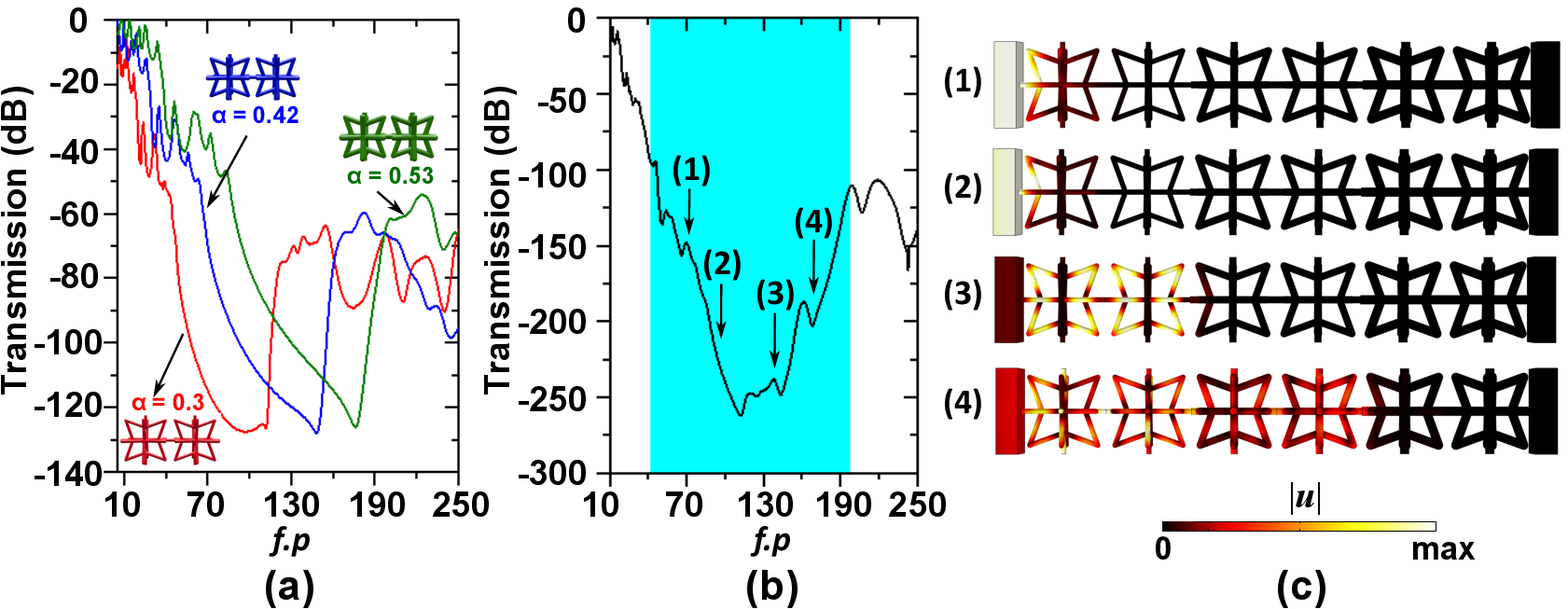}
\caption{\label{FourthFigure}(a) Transmission coefficient through two-unit cells for the cases of $\alpha = 0.3$, $\alpha = 0.42$ and $\alpha = 0.53$ (b) Elastic wave transmission in the $\Gamma$X direction, through the hybrid sample made up of the three sets of units cells (c) Displacement field amplitude through the sample for the four different frequencies indicated in the transmission spectrum}
\end{figure*}

Additionally, the local resonance bandgaps can also be tuned via the geometrical parameters of the unit cell. To understand this tailorability, the evolution of the bandgaps along the frequency spectrum was examined as a function of the geometrical parameters $\theta$ and $\alpha$. Figures \ref{ThirdFigure}(a) and (b) show complete bandgaps once again represented by the shaded cyan regions as a function of $\theta$ and $\alpha$ respectively. In the case of Fig. \ref{ThirdFigure}(a), the value of $\alpha$ if fixed to be 0.3 and the results thus indicate that for a constant strut diameter, multiple bandgaps can be opened and closed upon tuning $\theta$. Starting from $\theta$ = 4$^{o}$, the structure displays a wide bandgap centered around the normalized frequency, $f.p$ = 82, which continues to exist for higher values of $\theta$ as well. This bandgap reaches its maximum at $\theta$ = 20$^\circ$ and it can be seen by means of the dashed blue line that its central frequency is almost constant until this point. For $\theta > 20^\circ$, however, the bandgap width and central frequency decrease until it completely closes for $\theta$ = 40$^\circ$. Alongside this wide bandgap, multiple smaller-width gaps exist for different values of $\theta$. A high frequency bandgap occurs around the normalized frequency, $f.p$ = 116 for $\theta < 12^\circ$ while a very low frequency one exists at $f.p$ = 30 for $\theta \geq 12^\circ$. In addition to these, a second-high frequency gap emerges for $\theta > 20^\circ$ with a center frequency that rapidly decreases when increasing $\theta$. While $\theta$ approaches 28$^\circ$ however, the low frequency bandgap (located below $f.p$ = 30) reaches its maximum width of 23.6\% and then closes when $\theta$ = 40$^circ$. Simultaneously though, a second low frequency band begins from $\theta > 28^\circ$. Likewise, Fig. \ref{ThirdFigure}(b), shows the effect of varying rod diameters on the bandgaps when the angle $\theta$ remains fixed at 20$^\circ$ (this corresponds to the widest bandgap that was seen in the Fig. \ref{ThirdFigure}(a)). As can be seen here, the structure displays two complete bandgaps, both of which have their central frequencies that increase almost linearly with the increase of the rod diameter. The width of the lowest bandgap increases slightly to reach a maximum of 18.2\% at $\alpha$ = 0.6 and then decreases with further increase of $\alpha$. Also, the width of highest bandgap decreases with the increase of the diameter: from 82.8\% for $\alpha$ = 0.3 to 50.3\% for $\alpha$ = 0.8. The widest bandgap for this structure is therefore noted to occur when $\alpha $= 0.3 and $\theta$ = 20$^\circ$, which was hence the candidate for the analysis in the previous section. It is interesting to make note however, that a variety of bandgap widths and combinations can be made possible by simply tuning the geometrical parameters of the proposed design. 

The ease of  tailorability and extremely high wave attenuation was then leveraged to put forward a hybrid structure that is thin and lightweight yet equipped with an ultra-wide bandgap. This was made possible by combining three pairs of unit cells with different but overlapping bandgaps. Here, the three unit cells chosen had the same $\theta$ value but different $\alpha$ values. The values for $\alpha$ (i.e. the strut diameters ‘$d$’) were chosen from the bandgap analysis in Fig. \ref{ThirdFigure}(b). At first, the transmission performance for each of the cases, were examined when only two-unit cells were considered. Figure \ref{FourthFigure}(a) here, shows the transmission results for $\alpha$ = 0.3, 0.42 and 0.53 which are denoted by the red, blue and green curves respectively. High attenuation of the wave is clearly seen in the desired frequency ranges for each of the cases: $f.p$ from 46 to 111 for $\alpha$= 0.3; from 68.2 to 150.8 for $\alpha$ = 0.42 and from 90 to 184 for $\alpha = 0.53$. The three units were then combined and the wave transmission through this hybrid design can be seen in Fig. \ref{FourthFigure}(b). An overall attenuation can be seen over the frequency ranges of the three individual bandgaps seen in Fig. \ref{FourthFigure}(a). Figure \ref{FourthFigure}(c) shows the amplitude of the total displacement field for the hybrid micro-lattice along its $\Gamma$X direction, at the frequencies indicated by (1) inside the first bandgap but outside the second, (2) inside the common regions of the three bandgaps, (3) inside the second and the third bandgap and (4) inside the third bandgap but outside the others. The waves of frequency (1) and (2) are completely attenuated by the first two units as these are inside the bandgap of the first two-unit cells (which correspond to $\alpha$ = 0.3). Higher attenuation is seen in (2) since the wave at this frequency does not propagate through any of the units. In the case of (3), the wave propagates through the first two units and then gets completely attenuated by the units corresponding to $\alpha$ = 0.42 and $\alpha$ = 0.53, since its frequency lies in the overlapping region of the bandgaps of $\alpha$ = 0.42 and 0.53. Finally, for the frequency indicated by (4), the wave propagates through the first 4 units since it is no longer in the bandgap region associated to $\alpha = 0.3$ and $0.42$, but it gets completely reflected by the two units that correspond to $\alpha$ = 0.53. This hybrid sample would thus only be of 6.5 cm in thickness, but will facilitate an ultra-wide bandgap of width 130\%. 

In summary, we have introduced a micro-lattice based metamaterial, that is equipped with the unique capability of attenuating elastic waves in all directions and over a wide frequency range, alongside possessing the second order functionality of exhibiting a negative Poisson's ratio. Its fabrication can be made possibly by projection micro-sterelithography \cite{Zheng2016,Zheng2014,Gerard2019} and the experiments can be performed by a measurement platform similar to the transmission simulations performed here. The performed simulations thus account well for a future experimental realization of this metamaterial. Furthermore, the tunability that is facilitated by this design, can be deployed for reconfigurable metamaterials for active vibration control. The proposed metamaterial could thus pave way toward the development of a variety of futuristic wave based devices.

\end{document}